\documentclass[pre,aps,reprint,superscriptaddress,floatflix,twocolumn,amsfonts,amsmath]{revtex4-1}
\usepackage{graphicx}
\usepackage{subfigure}
\usepackage{amssymb}
\usepackage{latexsym}
\usepackage{soul}
\usepackage{color}

\newcommand{\stl}[1]{\mbox{$ \hspace{0.1em}
      \stackrel{\rule{0.4pt}{0.275ex}\hspace{0.40em} \!\!\!
      \overline{\hspace{0.06em}\vphantom{\rule{0.4pt}{0.0ex}}
      \hphantom{\mbox{$\displaystyle #1$}}
      \hspace{0.06em}  } \!\!\!\hspace{0.40em}\rule{0.4pt}{0.275ex}}
      {#1}\hspace{0.2em}$}}

\begin{document}
\title{Stochastic kinetics reveal imperative role of anisotropic interfacial tension to 
determine morphology and evolution of nucleated droplets in nematogenic films} 
\date{\today}
\def\iisc{\affiliation{Centre for Condensed Matter Theory, Department of Physics, Indian Institute
of Science, Bangalore 560064, India}}
\author{Amit Kumar Bhattacharjee} \iisc
\email{Email address: amitb@physics.iisc.ernet.in}

\begin{abstract}
For isotropic fluids, classical nucleation theory predicts the nucleation rate, barrier 
height and critical droplet size by accounting for the competition between bulk energy 
and interfacial tension. The nucleation process in liquid crystals is less understood. 
We numerically investigate nucleation in monolayered nematogenic films using a mesoscopic 
framework, in particular, we study the morphology and kinetic pathway in spontaneous 
formation and growth of droplets of the stable phase in the metastable background. The 
parameter $\kappa$ that quantifies the anisotropic elastic energy plays a central role 
in determining the geometric structure of the droplets. Noncircular nematic droplets 
with homogeneous director orientation are nucleated in a background of supercooled 
isotropic phase for small $\kappa$. For large $\kappa$, noncircular droplets with integer 
topological charge, accompanied by a biaxial ring at the outer surface, are nucleated. 
The isotropic droplet shape in a superheated nematic background is found to depend on 
$\kappa$ in a similar way. Identical growth laws are found in the two cases, although 
an unusual two-stage mechanism is observed in the nucleation of isotropic droplets. 
Temporal distributions of successive events indicate the relevance of long-ranged 
elasticity-mediated interactions within the isotropic domains. Implications for a 
theoretical description of nucleation in anisotropic fluids are discussed.
\end{abstract}

\maketitle

\section*{Introduction}
\label{sec:1}

A fluid exhibiting a first order phase transition can transit from an unstable to a stable 
phase through spinodal decomposition and coarsening, where irregular domains of the stable 
phase emerge spontaneously and combine to minimize the surface energy. In contrast, 
transformations from a metastable state occur via nucleation and growth in which droplets 
of the stable phase are formed in the metastable state and these droplets grow and coalesce 
to increase the fraction of the stable phase in the system. A classic example of this 
phenomenon is supercooled water freezing into ice via nucleation and growth\cite{fahren}. 
Nucleation in solid solutions is followed by Ostwald ripening\cite{voor}, while metallic 
alloys and bulk metallic glasses conventionally display dendritic growth due to anisotropic 
surface effects\cite{liglkurz}.  

Many fundamental problems in surface interfacial science are concerned with the morphology 
of the nucleated phase, its growth rate, the first passage time as well as the kinetic route 
to equilibrium. Questions about droplet morphology are especially pertinent in studies of 
nematogenic fluids, where the anisotropy associated with the tensorial structure of the 
order parameter is one of the important factors in the description of the nucleation 
process\cite{cutodijk,kowamcvi}. The microstructure of the nucleus is determined by a 
nontrivial interplay of competing energies: (i) the anisotropic elastic energy associated 
with deformations of the tensorial order in the bulk, (ii) the anisotropic interfacial 
tension related to the director anchoring at the interface between the two phases, and 
(iii) any external forcing that may be present, e.g. equilibrium thermal fluctuations. 
Thus, aspherical shape of droplets, complex growth law etc. are to be expected and the 
nucleation rate may itself lack a precise definition\cite{absorey}. 

Recently, liquid crystalline phases have found a multitude of applications in nanoscience\cite{bisan}. 
Droplet shapes play a crucial role in ink-jet technology\cite{altakhyan}, switching and 
bistable devices\cite{tslehokw}, photovoltaics as well as in bio-sensor applications with 
living liquid crystals\cite{zhsolaara}. Early experiments found evidence for aspherical 
spindle-shaped droplets called tactoids\cite{bernfank}. Such nuclei were later obtained
in theoretical studies assuming homogeneous director distribution inside the droplet\cite{herring,
chandrasekhar,virga}. Progress was hindered for several decades because experimental 
characterization of early-stage supercritical droplets was not possible. Recently, long 
carbon nanotubes have been used in optical microscopy to characterize nematic tactoids\cite{jabeselesmscpas}. 

Computer simulations have traditionally played an important role in the development of 
an understanding of the kinetics of nucleation and growth. Computer simulations of nucleation 
processes have to address problems in defining the droplets unambiguously and in developing 
algorithms to sample rare events. Monte Carlo (MC) studies of hard spherocylinders have been 
performed, where ellipsoidal clusters with homogeneous director orientation are 
nucleated\cite{cutodijk,cutoroijdijk}. More recently, spherical nanodroplets with a radial 
hedgehog defect, accompanied by a Saturn-ring at the core and bipolar pole-centered boojum 
defects with uniform field structure have been reported\cite{pehuguorpab}. It is worth 
mentioning that kinetic pathways in MC simulations can be misleading, as the algorithm 
samples the Gibbs distribution in equilibrium without obeying the natural dynamics of 
the system. Slower growth following a diffusive kinetics are reported in molecular dynamics 
(MD) simulation and experiments\cite{brakrazum,chehamshe}. Recent studies have examined 
the morphology of freely suspended aspherical nanodroplets\cite{varibezan}. Although MD 
provides a comparatively well-defined temporal evolution than MC, nematic ordering is 
often best discussed using coarse-grained methods for which a top-down approach works 
very well\cite{varibezan} due to the scale invariance of the dynamical equations, allowing 
its applicability from astrophysical scales, {\it e.g.} the Kibble-Zurek mechanism\cite{kibble,zurek}, 
down to nanoscales. We use this approach in our work.

Nematic order is described by a symmetric, traceless tensor ${\bf Q}$, which in component 
form reads \cite{klelav} $Q_{\alpha\beta} = [S(3n_{\alpha}n_{\beta}-\delta_{\alpha\beta}) + 
B_2(l_{\alpha}l_{\beta} - m_{\alpha} m_{\beta})]/2$, where $(\alpha,\beta) \equiv (x, y, z)$ 
denote the Cartesian directions in a local frame of reference with $S = \langle {\cos^{2}\theta - 
1/3}\rangle$ and $B_2=\langle {\sin^{2}\theta\cos2\phi}\rangle$ the scalar degree of uniaxial 
and biaxial order, respectively, ($\theta,\phi$) the polar and azimuthal angles and averaging 
is done over a sufficient large coarse-graining volume. [${\bf n,l,m}$] denote the director, 
codirector and secondary director forming an orthonormal triad. The Ginzburg-Landau-de Gennes 
(GLdG) free energy consists of a homogeneous bulk term and an elastic term representing the 
free-energy cost of distortions due to inhomogeneity, namely $\mathcal{F} = \mathcal{F}_{bulk} + 
\mathcal{F}_{elastic}$, where
\begin{eqnarray}
\label{eq:frenerQ}
\mathcal{F}_{bulk} &=& \int d^3{\bf x} \Big[\frac{1}{2}A\mathrm{Tr}{\bf Q}^{2} + 
\frac{1}{3}B\mathrm{Tr}{\bf Q}^{3} + \frac{1}{4}C(\mathrm{Tr}{\bf Q}^{2})^{2}\Big], \nonumber \\ 
\mathcal{F}_{elastic} &=& \int d^3{\bf x} \Big[\frac{1}{2}L_{1}(\partial_{\alpha}
Q_{\beta\gamma})(\partial_{\alpha}Q_{\beta\gamma}) + \frac{1}{2}L_{2}(\partial_{\alpha}
Q_{\alpha\beta})\times \nonumber \\ 
&& (\partial_{\gamma}Q_{\beta\gamma}) + \frac{1}{2}L_{3}Q_{\alpha\beta}(\partial_{\alpha}
Q_{\gamma\delta})(\partial_{\beta}Q_{\gamma\delta})\Big].
\end{eqnarray}
\begin{figure}[ht]
\centering
\includegraphics[width=0.5\textwidth, height=0.24\textwidth]{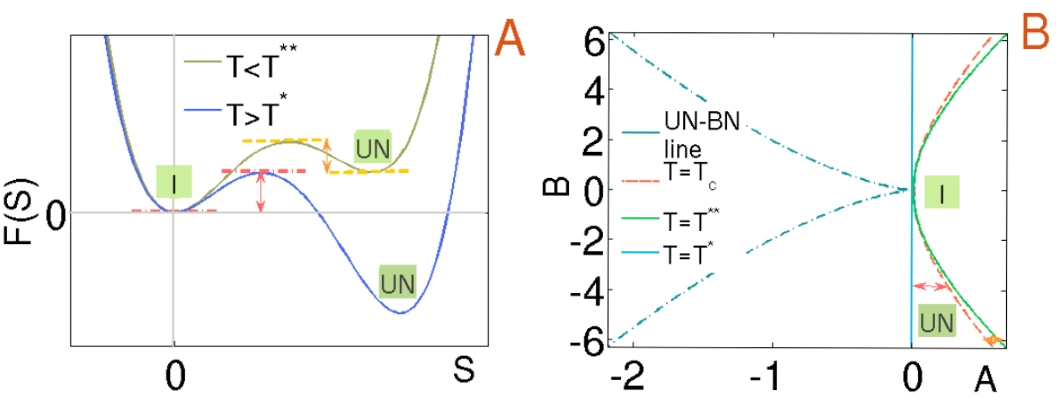}
\caption{\label{fig:schematic}(Color Online) (A) Schematic illustration of the free energy 
with scalar order and (B) corresponding phase diagram with stable and metastable states. 
[I] and [UN] denote the isotropic and uniaxial nematic minima. Second order uniaxial-biaxial 
[UN-BN] line is also shown and the barrier height is marked in red (orange) for supercooling 
(superheating), with the spinodal temperatures\cite{degenpro,bhmeadd} denoted by $T^*$, $T^{**}$. 
Recall that $T^*$, $T^{**}$ and the clearing temperature $T_c$ correspond to $A=0, B^2/24C$ 
and $B^2/27C$ respectively\cite{pcl}. For example in 5CB, $T^*=34.2^{\circ}$C, 
$T^{**}=34.47^{\circ}$C and $T_{c}=34.44^{\circ}$C.}
\end{figure}

$\mathcal{F}_{bulk}$ is displayed in fig.[\ref{fig:schematic}(A)] that exhibits an asymmetric 
well landscape characterizing the weakly first order nature of the isotropic-nematic phase 
transition. The phase diagram in fig.[\ref{fig:schematic}(B)] is derived from $\mathcal{F}_{bulk}$, 
where the temperature dependence is contained in the parameter $A=A_0(1-T/T^{*})$  and the 
parameter $B$ depends on the size disparity\cite{gralondej}. Minimizing $\mathcal{F}_{bulk}$ 
with respect to $S$ yields the equilibrium value 
\begin{equation} 
S_{eq} = -B/ 6C + \sqrt{B^2/36C^2- 2A/3C}
\end{equation} 
with the clearing point value $S_c=-2B/9C$. 
\begin{table*}
\small
\begin{tabular*}{1.0\textwidth}{@{\extracolsep{\fill}}cccccccccc}
\hline
Fig. & $\Gamma (Poise^{-1})$ & $A (Jcm^{-3})$ & $B(Jcm^{-3})$ & $C(Jcm^{-3})$ & $L_1 (10^{-7}dyn)$ & $\kappa$ & $\lambda (\mu m)$ \\
\hline
\hline
(2,3A-3D) & $1$ & $10^{-3}$ & $-0.5$ & $2.67$ & $(0.025,0.012,0.012,0.01)$ & $(-1,0,1,18)$ & $(3.38,2.56,3.31,8.44)$ \\
(3E-3G,4-5)  & $1$ & $10^{-3}$ & $-0.5$ & $2.67$ & $0.01$                     & $(-1,0,1,18)$ & $(2.14,2.34,3.02,8.44)$ \\
\hline
(6,7A-7D) & $1.25\times10^{-2}$ & $0.38019$ & $-4.0$ & $1.67$ & $(1.5,0.895,0.66,0.4)$ & $(-1,0,1,6)$ & $(3.95,3.35,3.71,5)$ \\
(7E-7G,8) & $1.25\times10^{-2}$ & $0.38019$ & $-4.0$ & $1.67$ & $0.4$                  & $(-1,0,1,6)$ & $(2.04,2.24,2.89,5)$ \\
\end{tabular*}
\begin{tabular*}{1.0\textwidth}{@{\extracolsep{\fill}}cccccc}
\hline
Fig. & $\Upsilon^{*}$ & Fig. & $\Upsilon^{*}$ & $k_BT (J)$ & $t^{*}$ \\
\hline
\hline
(2,3A-3D) & $(1.80,1.04,1.73,11.3)\times10^{-5}$ & (3E-3G,4-5) & $(2.89,8.66,14.4,113)\times10^{-6}$ & $2.0807\times10^{-7}$ & $2.6\times10^{-3}$ \\
\hline
(6,7A-7D) & $(1.77,1.27,1.56,2.83)\times10^{-1}$ & (7E-7G,8) & $(4.72,5.67,9.44,28.3)\times10^{-2}$ & $6\times10^{-3}$ & $3.33\times10^{-3}$\\
\hline 
\end{tabular*}
\caption{\label{tbl:GLdGparam} Values of parameters used to obtain the plots shown in fig.(2-8), where 
a box of size $L_x=L_y=96 \mu m$ with grid spacing $\Delta x = \Delta y=1 \mu m$ and time step 
$\Delta t=1 \mu s$ are considered. For 5CB, the data correspond to a temperature of $34.27^{\circ}$C 
in fig.(2-5) and $34.46^{\circ}$C in fig(6-8). Definition of the parameters are given in the Methods 
section.}
\end{table*}

The first two terms of $\mathcal{F}_{elastic}$ correspond  to isotropic and anisotropic 
elasticity with the final term being a higher order contribution. The elastic constants 
$L_1,L_2,L_3$ are obtained from experimental measures of Frank-Oseen splay ($K_1$), twist 
($K_2$) and bend ($K_3$) elastic constants via the relation \cite{schtrim}, 
\begin{eqnarray}
\label{eq:k1k2k3}
& K_1 = {9S^2}L_1\big(2 + \kappa - \Theta S\big)/4, K_2 = {9S^2}L_1\big(2 - \Theta S\big)/4, \nonumber \\ 
& K_3 = {9S^2}L_1\big(2 + \kappa + 2\Theta S\big)/4,
\end{eqnarray} 
where $\kappa = L_2/L_1$ and $\Theta = L_3/L_1$ ($L_1>0$). Third order terms can be 
neglected ($\Theta=0$) leading to degenerate splay and bend with twist either large or 
small depending on the sign of $\kappa$ \cite{wincurerey,klelav}. Thus the GLdG theory 
loses its validity if bend and splay constants are very different. The {\it one elastic 
constant} approximation is often considered for analytic convenience, where $K_1=K_2=K_3$ 
corresponds to $\kappa=0$. However, experimental measures of elastic constants in units 
of $10^{-7} dyn$ and GLdG coefficients in units of $Jcm^{-3}$ for (a) 5CB at 25$^{\circ}$C 
are $K_1=6.4, K_2=3, K_3=10, B=7.2, C=8.8$ and (b) MBBA at 25$^{\circ}$C are $K_1=6, K_2=4, 
K_3=7.5, B=2.66, C=2.76$\cite{blchig}. $\Theta=0$ gives for 5CB, $L_1 = 0.649, \kappa = 40.667$ 
and for MBBA, $L_1 = 8.6534, \kappa = 1.2$. This explains why the {\it one elastic approximation} 
is inappropriate in a description of certain nematogenic materials. 

Using this free energy, the geometric structure of monolayered droplets has been studied 
analytically in the past two decades, either making several simplifying assumptions\cite{lafrmai,sokihes,prinschoot}, 
or through an exact computation\cite{bhmeadd}. Going beyond the Frank-Oseen description 
of the elastic energy\cite{dauelast} and without enforcing any phenomenological Rapini-Papoular 
(RP) surface energy term\cite{rapinipap}, noncircular nematic droplets with integer topological 
charge have been found to grow ballistically\cite{lafrmai} in a deterministic (no thermal 
noise) calculation.

Homogeneous nucleation kinetics can not be studied in the deterministic GLdG framework 
because droplets of the stable phase cannot spontaneously nucleate in a metastable medium 
in the absence of thermal fluctuations. Near the transition point, droplet growth is governed 
by capillary forces rather than the small free energy difference or volume driving force, 
where fluctuations play a crucial role\cite{wincurerey1}. To understand how fluctuations 
influence the dynamics and microstructural evolution, one needs (i) the theoretical formulation 
of a stochastic GLdG description of the dynamics and (ii) a numerical prescription to integrate 
the stochastic equation for the orientation tensor\cite{stratv} paying special attention to the 
structure of the noise and satisfying the fluctuation-dissipation theorem (FDT). The first 
question was addressed by Stratonovich\cite{stratv} by writing an overdamped Langevin equation 
in model-A relaxational dynamics that excludes coupling to any external hydrodynamic flow as\cite{halhoh,bhmeads} 
\begin{equation}
\label{eq:Qdynamics}
\partial_t Q_{\alpha\beta} = - \Gamma\Big[\delta_{\alpha\mu}\delta_{\beta\nu} +
\delta_{\alpha\nu}\delta_{\beta\mu} - \frac{2}{3}\delta_{\alpha\beta}
\delta_{\mu\nu}\Big] {\delta \mathcal{F}\over\delta Q_{\mu\nu}} + 
\xi_{\alpha\beta},
\end{equation}
where  the coefficient of rotational diffusion $\Gamma$ controls the relaxation rate and the 
symmetric traceless tensorial random force ${\pmb\xi}$ satisfies the property, 
$\langle \xi_{\alpha\beta} \rangle = 0, \langle \xi_{\alpha\beta}({\bf x}, t) \xi_{\mu\nu}({
\bf x^{\prime}}, t^{\prime})\rangle = 2k_BT \Gamma[\delta_{\alpha\mu}\delta_{\beta\nu} + 
\delta_{\alpha\nu}\delta_{\beta\mu} - \frac{2}{3}\delta_{\alpha\beta} \delta_{\mu\nu}] 
\delta({\bf x - x^{\prime}})\delta(t - t^{\prime})$ to ensure FDT and thus Gibbs distribution 
at equilibrium \cite{daustat}. $k_B,T$ and brackets denote the Boltzmann constant, equilibrium 
temperature and average over the probability distribution of ${\pmb\xi}$. The first term in 
$\mathcal{F}_{elastic}$ leads to $L_1 {\partial^2\bf Q}$ in equation(\ref{eq:Qdynamics}), 
indicating an isotropic diffusion of ${\bf Q}$. The second term in $\mathcal{F}_{elastic}$ 
leads to $L_2\stl{\pmb\partial(\pmb\partial\cdot\bf Q)}$ in the evolution equation, resulting 
in an orientation dependent {\bf Q}-diffusion that leads to two diffusion constants in the 
nematic phase. The anisotropy is controlled by the parameter $\kappa$ defined above. An efficient 
method for numerical integration of this equation was developed in a recent work of the author 
\cite{bhmeads} that motivated the present study. 

Classical nucleation theory (CNT) estimates the critical size of a droplet, the barrier 
height and the nucleation rate using the assumption that nucleation proceeds via the formation 
and expansion of spherical droplets\cite{rayl,becdor}. The excess free energy of a droplet 
is obtained as $\Delta {\mathbb {F}} = -4\pi R^3 \rho_N \Delta\mu/3 + 4\pi R^2 \sigma$, 
where $R$ is the droplet radius, $\rho_N$ is the density of the nucleated phase, 
$\Delta\mu={\mathbb{L}\Delta T}/{T^{*}}$ is the chemical potential difference with $\mathbb{L}$ 
being the emitted latent heat due to a change in temperature $\Delta T$ and $\sigma$ is the 
interfacial surface tension. Maximizing $\Delta \mathbb{ F}$ with respect to $R$ yields 
$R_c = {2\sigma}/{\rho_N |\Delta\mu|}$ and the barrier height $\mathbb{F}_c = {16\pi \sigma^3}/{3 
\rho^2_N (\Delta\mu)^2}$. The nucleation rate is defined as $I = \mathcal{A} e^{-\mathbb{F}_c/k_BT}$ 
where $\mathcal{A}$ is a kinetic prefactor often hard to measure in experiments, making the rate 
calculation a formidable problem. 

For droplets formed in a nematogenic material, due to the inherent anisotropy in the field 
variables, the free energy takes the form 
\begin{equation}
\label{eq:Fcntnem}
\Delta \mathbb{ F} = -\int_V d^3x [\rho_N \Delta \mu ({\bf Q})] + \int_{{\pmb\partial} \mathcal{S}} 
d^2x [\sigma({\pmb\partial\bf Q})],
\end{equation} 
where $V$ and ${\pmb\partial}\mathcal{S}$ respectively denote the transformed volume and 
the enclosing surface. The complexity that renders an analytical insight difficult lies 
in the nontrivial coupling between principal values and principal axes of the {\bf Q}-tensor. 
For an ellipsoidal droplet with homogeneous director distribution, analytic expressions can 
be derived from the above equation \cite{cutoroijdijk,popaslu} without considering a RP-term. 
However, for a noncircular droplet with an embedded defect, singular volume and surface 
integrals restrict the applicability of an analytic approach. The interfacial surface 
tension is thermodynamically defined as the excess surface energy per unit area. The first 
and second terms in $\mathcal{F}_{elastic}$ contribute to the isotropic and anisotropic parts 
of the surface energy, respectively. The excess anisotropic surface energy is controlled by 
the parameter $\kappa$ defined earlier, that differentiates between strong and weak anchoring 
of the director at the interface. 
\begin{figure*}[ht]
\begin{centering}
\includegraphics[width=\textwidth, height=0.68\textwidth]{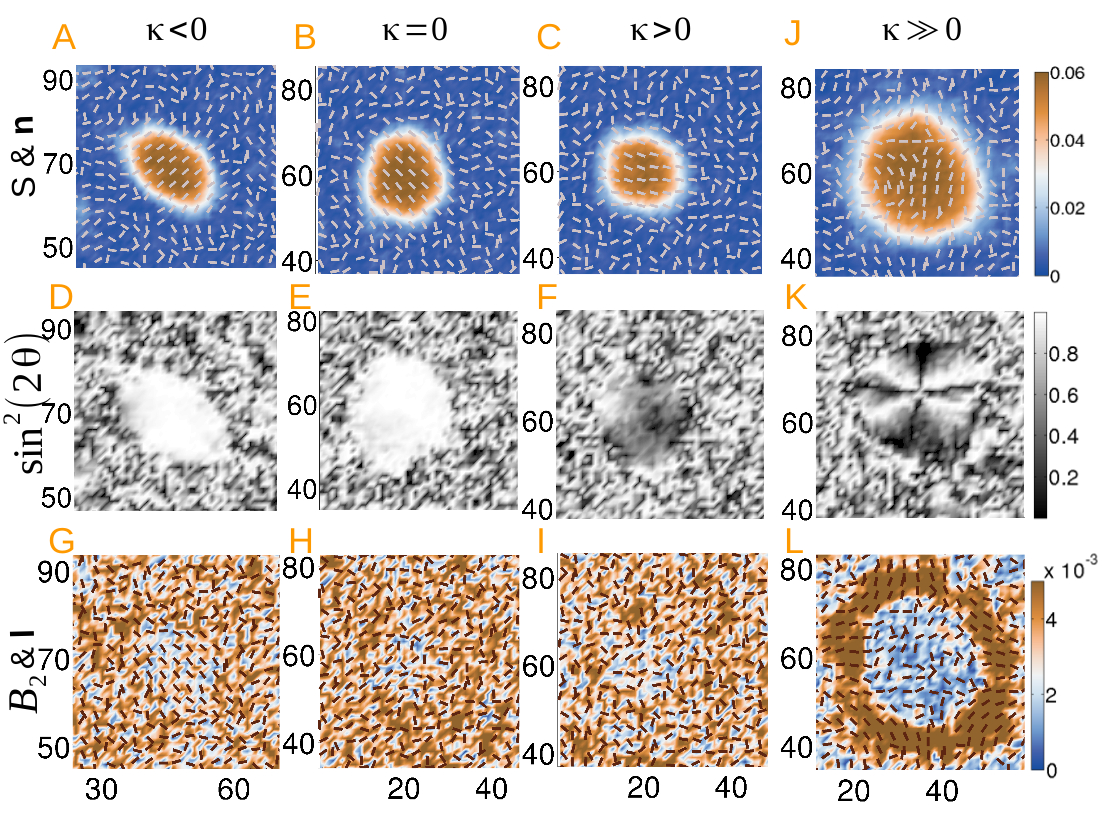}
\par\end{centering}
\centering{}\caption{\label{fig:INmorph} (Color Online) Nematic droplet structure in terms 
of the uniaxial order parameter and director orientation for $\kappa=-1$ at $t=4031\tau$ (panel A), 
$\kappa=0$ at $t=4623\tau$ (panel B), $\kappa=1$ at $t=6083\tau$ (panel C) and $\kappa=18$ 
at $t=8099\tau$ (panel J) in the post-nucleation stage of the kinetics. Panels (D-F, K) display 
the corresponding Schlieren texture which is proportional to $sin^2(2\theta)$ and panels (G-I, L) 
depict the degree of biaxiality and the codirector orientation. The critical radius for $L_1=0.01$ 
and $\kappa = (-1,0,1,3,6,18)$ turns out to be $R_c = (7.72, 9.35, 11.05, 11.8, 4.4, 0.49)$. Scalar 
field values are rendered in false colour.}
\end{figure*}
 
Nucleation and growth are often characterized by the Johnson-Mehl-Avrami-Kolmogorov (JMAK) 
equation \cite{avram1,avram2,avram3} $x(t) = 1 - e^{(t/\mathcal{T})^m}$, where $x(t)$ is 
the volume fraction of the nucleated phase, $m$ depends on the shape of the droplet and 
$\mathcal{T}$ is a constant related to the growth velocity $v$. For isolated spherical 
droplets with number density $n$, simple analysis shows that $m=3, \mathcal{T}=(3/4\pi nv^3)^{1/3}$. 
However, if we consider expanding ellipsoidal droplets where the long and the short axes increase 
self-similarly, then the parameters turn out to be $m=3, \mathcal{T}=(9/8\pi nv^3)^{1/3}$. 
Higher exponents and fractional exponents are also seen in experiments and conventionally 
calculated through a plot of 
\begin{equation}
\label{eq:JMAK}
ln[-ln\{1-x(t)\}] =  mln(t) - mln \mathcal{T}
\end{equation}
{\it versus} $ln(t)$. While the exponent $m$ is dictated by the dimensionality of the droplet, 
a departure from the predicted value suggests the inapplicability of simple theory and breakdown 
of CNT. In our results, droplet represents a quasi two-dimensional ``raft''-like geometry 
formed in monolayered film. 

\section*{Results}
\label{sec:3}
We first discuss the tensorial microstructure and evolution of thermally generated nematic 
droplets in a supercooled isotropic phase. This is done for varying anisotropic surface energy 
and the results are compared with the predictions of classical theories of nucleation to test 
their applicability. We probe the role of long range elasticity mediated interaction on the 
distribution of the first passage time between successive events. We also consider the 
nucleation of isotropic droplets in a superheated nematic phase. The numerical values of the 
parameters used in our simulations are tabulated in Table~\ref{tbl:GLdGparam}. 
\begin{figure*}[ht]
\includegraphics[width=1.0\textwidth, height=0.65\textwidth]{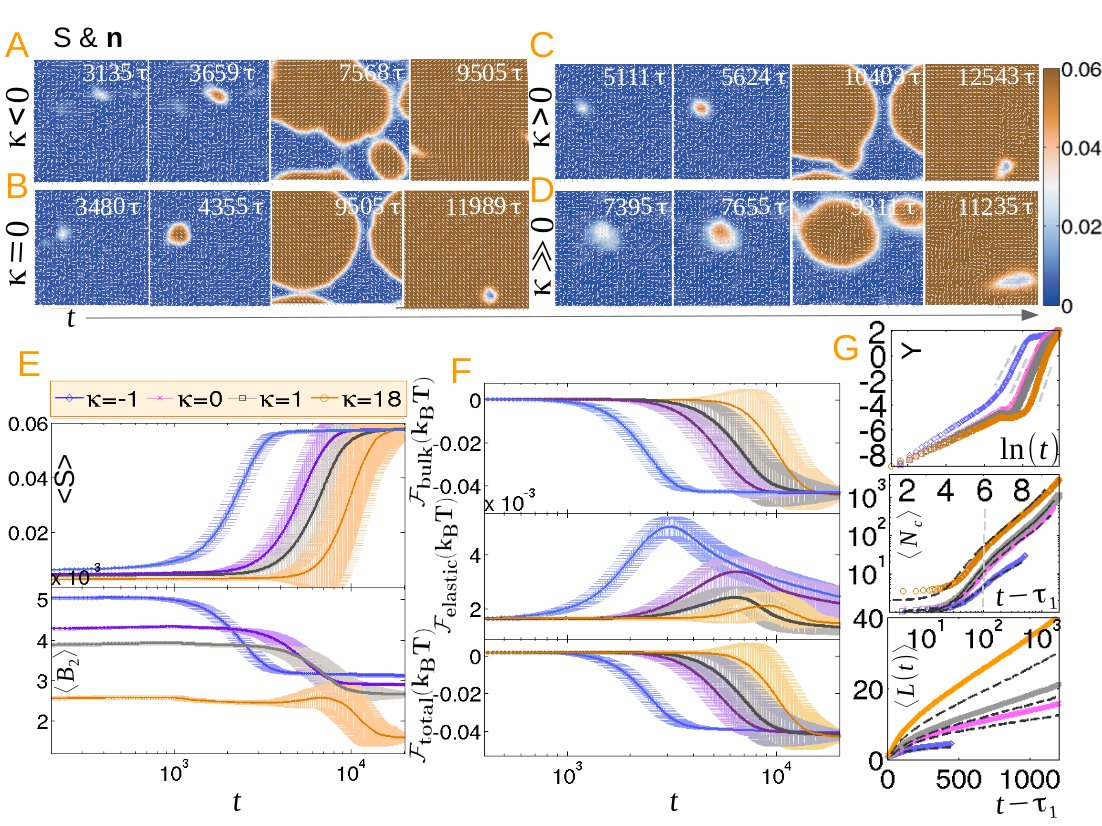}
\caption{\label{fig:FSTevolIN} (Color Online) Panel(A-D): Evolution of the scalar uniaxial order 
and director structure at pre, post, intermediate and late stages of the kinetics for higher 
surface energy and different $\kappa$ (See Supplementary Animations S1 and S2). Panel (E) displays 
the average uniaxial order $\langle S\rangle$ and biaxial order $\langle B_2\rangle$ while 
panel (F) shows the bulk, elastic and total free energy of the film. Plots of the JMAK eq.(\ref{eq:JMAK}) 
are displayed in the upper panel (G) with exponents $m=(3.145,3.425,3.465,3.98)$ for $\kappa=(-1,
0,1,18)$ in ascending order. Finally, evolution of the number of points in a tagged cluster 
(coloured symbols) as well as the average cluster size (black dotted lines) for different $\kappa$ are 
shown in the middle panel (G), while the lower panel (G) displays the evolution of the length scale 
obtained from the middle panel (G). $800$ independent realizations are sampled to 
obtain the graphics in panel(E-G).} 
\end{figure*}

\subsection*{Nucleation in supercooled isotropic phase}
Our central findings are summarized in terms of the droplet morphology, evolution of the 
{\bf Q}-tensor and the free energy, growth law and temporal distribution of nucleation events. 
Fig.(\ref{fig:INmorph}) shows the supercritical droplet structure at the post-nucleation stage 
in terms of the uniaxial order $S$ and the director distribution {\bf n}, the biaxial order $B_2$ 
and codirector distribution ${\bf l}$, as well as the Schlieren texture for different values of 
$\kappa$ chosen to ensure the positivity of the Frank elastic constants. The nucleated droplet 
in panel (B) is circular in the {\it one elastic constant} approximation ($\kappa=0$) while the 
droplet in the weak anchoring limit (small $\kappa$) shown in panels (A,C) is noncircular. As 
indicated by the orientation of {\bf n}, homeotropic anchoring at the interface is preferred 
for $\kappa=-1$, corresponding to $K_2=2K_1$ (defined in eq.(\ref{eq:k1k2k3})), where the 
uniform director inside the droplet orients perpendicular to the long axis. For negative 
values of $\kappa$, its magnitude cannot be arbitrarily large as an unphysical correlation 
length is numerically unavoidable for the analytical lower bound $\kappa>-6$\cite{degennes}. 
On the other hand, planar anchoring is favoured for $\kappa=1$ corresponding to $K_2=2K_1/3$, 
where the director orients parallel to the long axis. For a flat interface, the total energy 
is lowered for planar or homeotropic director anchoring for $K_2$ being smaller or larger than 
$K_1$. This result is often termed as the {\it de Gennes ansatz}\cite{degennes}. Though this 
ansatz does not hold for curved interfaces (shown in the Supplementary Information), our 
results agree reasonably well with it. This result is also in agreement with deterministic 
GLdG calculations for bubbles created by hand\cite{bhmeadd} and MC, MD simulations\cite{cutodijk,varibezan}. 
This result, however, contradicts those of Ref.~\cite{lafrmai}, where encapsulated 
integer-charged defects are reported inside an artificially constructed droplet for $\kappa$ 
in the range $(-4/7,4/3)$ [parameter $K$ in this study is related to $\kappa$ by $\kappa=2K/(1-K)$]. 
Uniformly white textured domains in panels (D-F) are indicative of the homogeneous director 
distribution in panels (A-C). Finally, panels (G-I) illustrate that $B_2$ has a small value 
(the order is uniaxial without any codirector or secondary director ordering) except for 
$\kappa=-1$. Biaxial fluctuations are visible in the isotropic film as $\langle B_2^2\rangle\neq0$ 
due to the presence of stochastic forcing.
 
This picture, however, changes dramatically in the strong anchoring limit ($\kappa\gg0$) as 
evident in the panels (J-L), where the microstructure at $\kappa=18$, corresponding to 
$K_2=K_1/10$, is depicted. Nonuniform director orientation inside the noncircular droplet 
corresponds to four-brush texturing that represents a hyperbolic hedgehog defect. The 
topological charge of $-1$ is quantified through a Volterra process\cite{klelav}. This reveals 
that there exists a threshold value of $\kappa\gg0$, for which the surface anisotropy is large 
enough to distort the field structure inside the droplet to encapsulate a defect. While the 
generation and growth of a supercritical nucleus depends on the competition between bulk and 
surface contributions, with the latter increasing with $\kappa$, the shape and director 
configuration inside the nematic region strongly depend on the surface interfacial anisotropy. 
The codirector ${\bf l}$ and the secondary director ${\bf m}$ (not shown) also have a singular 
structure with $B_2$ reaching a maximum on a noncircular ring embedded in the outer region of 
the droplet. This is consistent with the understanding that a planar interface exhibits local 
biaxiality for large $\kappa$\cite{kabhadmen}. When approximating the droplets to be circular, 
the critical droplet size can be estimated in terms of the parameters in the GLdG free energy. 
As mentioned in the caption of fig.(\ref{fig:INmorph}), unreasonable values of $R_c$ are 
obtained as the droplets become more noncircular with a nonuniform director arrangement. 
However, no analytic formula for $R_c$ can be obtained within a stochastic GLdG theory.

Next we address the various stages of the kinetics. Panels (A-D) of fig.(\ref{fig:FSTevolIN}) 
illustrate the pre, post, intermediate and late stage structure of $S$ and ${\bf n}$ for 
different $\kappa$ and with large $L_1$, implying droplets with a large surface energy. 
Increasing the barrier height results in a prolonged pre-nucleation stage and fewer 
supercritical droplets emerge in the post-nucleation period. Droplets grow self-similarly, 
coalesce at the intermediate stage and span the system at the late stage without forming 
any defect-antidefect pair. However, for smaller surface energy and $\kappa\le0$, half-integer 
defects with two-brush textures emerge due to the coalescence of droplets that resembles 
a reduced uniaxial order within the defect core (see Supplementary Animation S1). The 
ordering kinetics proceeds via the annihilation of defects, thus reducing the total 
free energy of the film. For $\kappa=1$, structures similar to boojum defects emerge at 
opposite poles of the droplet, where $S$ has saturated to the equilibrium value without 
displaying any half-integer defects. For $\kappa=18$, the nematic region gradually 
encroaches the isotropic domain with $S$ saturating relatively quickly as compared to 
the integer defect annihilation kinetics. The four-brush texturing persists even at a 
very late stage without generating any two-brush texturing (see Supplementary Animation S2). 
  
To understand the role of $\kappa$ in the kinetics, the growth and decay of average 
uniaxial and biaxial ordering for small $L_1$ are depicted in panel (E). The sigmoidal 
profile of $\langle S\rangle$ in the upper panel with higher intermediate slope inbetween 
two smaller slopes at early and late stages of the kinetics is a typical characteristic of 
nucleation followed by a growth process. For $\kappa=-1$, $t<10^3$ is identified as the 
pre-nucleation stage where subcritical nuclei shrink to zero, while $t>10^3$ denote the 
emergence of the supercritical nucleus and growth by agglomeration. Finally, the saturation 
of $\langle S\rangle$ for $t>3\times10^3$ corresponds to the defect annealing process. As 
anticipated, the nucleation time is prolonged for increasing $\kappa$, resulting from 
increased surface energy and hence a higher barrier height. Thus the number of droplets 
decreases for higher surface anisotropy. The fraction of the stable phase, $x(t) (0<x<1)$ 
and the function $Y = ln[-ln\{1-x(t)\}]$ are computed from the profile of $\langle S\rangle$ 
and fits to the JMAK equation(\ref{eq:JMAK}) are displayed in the upper panel of (G). The 
intermediate slope indicated by the dashed grey lines with scaling exponent $m>2$ indicates 
a breakdown of the simple theory and inapplicability of a CNT description. $\langle B_2\rangle$ 
in the lower panel of (E) decreases in a step fashion as the nematic phase is approached. A 
nonzero biaxiality at equilibrium conveys a departure from a purely uniaxial nematic film, 
with the magnitude of $\langle B_2\rangle$  decreasing with increasing $\kappa$. $\langle B_2\rangle$ 
attains an intermediate maximum before a step decrease for $\kappa\gg0$. This is related 
to the coalescence of the biaxial rings shown in fig.[\ref{fig:INmorph}(L)]. 

The total (free) energy of the film and the contributions from bulk and elastic energies 
are displayed in the panel (F). $\mathcal{F}_{total}$ decreases monotonically with time.  
$\mathcal{F}_{elastic}$ is smaller than $\mathcal{F}_{bulk}$ by about an order of magnitude.
The elastic energy slowly increases and exhibits an overshoot before decreasing to attain 
the equilibrium value. The overshoot is maximized for $\kappa=-1$, arising from the coalescence 
of homeotropically anchored nematic droplets leading to a maximum in elastic energy. The 
overshoot gradually decreases with increasing $\kappa$, as less elastic energy is needed 
in combining planar anchored droplets. 
\begin{figure}[ht]
\centering
\includegraphics[width=0.35\textwidth, height=0.3\textwidth]{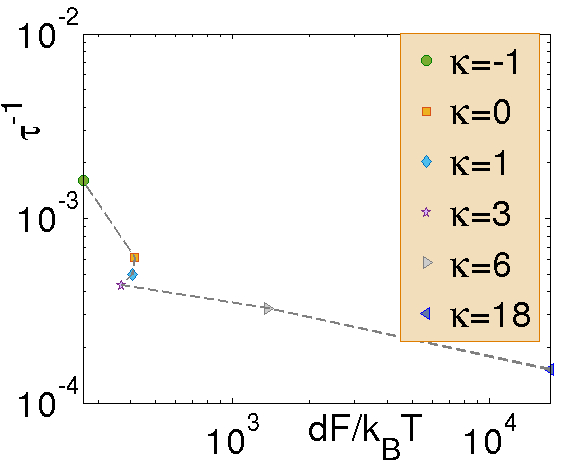}
\caption{\label{fig:IN_CNT} (Color Online) Nucleation rate as a function of barrier height 
for various $\kappa$. $400$ realizations for each $\kappa$ are sampled to obtain the graph.} 
\end{figure}

The growth of the first nucleated cluster $\langle N_c\rangle$ and the average cluster 
size at the post-nucleation stage before coalescence are displayed in the middle panel 
of (G). The growth law follows a polynomial form $\langle N_c \rangle = at^2 + bt +c$, 
where $a,b,c$ are fit parameters. As $\langle N_c \rangle$ scales as the square of the 
characteristic length $L$, the growth law for a tagged cluster is predicted to be
\begin{equation}
\label{eq:growth}
L(t) \sim (at^2+bt+c)^{1/2}.
\end{equation}
Evolution of the length scale for a tagged cluster, along with the average cluster size, 
is plotted in the lower panel of (G). In a brief period of the post-nucleation stage, 
the $at^2$ term in eq.(\ref{eq:growth}) can be neglected to obtain a diffusive, thermally 
limited regime where curvature elasticity and capillary forces play a more significant 
role than the free energy difference or the volume driving forces. Furthermore, the Laplace 
pressure is large due to a small radius of curvature and the surface interfacial tension, 
as well as the noncircular morphology of the droplet, induce local shear 
effects\cite{wincurerey,wincurerey1,huifas,evpusascroij}. A crossover to a ballistic volume 
driven growth regime at a later stage, marked with a grey vertical line in the middle panel, 
where the $bt$ term in eq.(\ref{eq:growth}) can be neglected, corresponds to a propagating 
interface front before droplet coalescence. The late stage ballistic growth in deterministic 
spinodal kinetics in confined circular films has been addressed earlier with a crystal growth 
equation supplemented to the deterministic GLdG framework\cite{wincurerey1}. Experiments in 
confined geometry, however, find diffusive dynamics at long times, which is incorporated in 
the deterministic GLdG formalism along with the equation for latent heat at the interface. 
As the heated interfacial temperature becomes comparable to that of the nematic bulk, growth 
reaches a diffusive steady state with an equal rate of generation and diffusion of latent 
heat\cite{chehamshe,absorey}. However, when the film is not confined, the latent heat effects 
are unimportant due to faster expulsion of heat from the droplet surface, leading to a long 
time ballistic growth. 
\begin{figure}[ht]
\centering
\includegraphics[width=0.5\textwidth, height=0.4\textwidth]{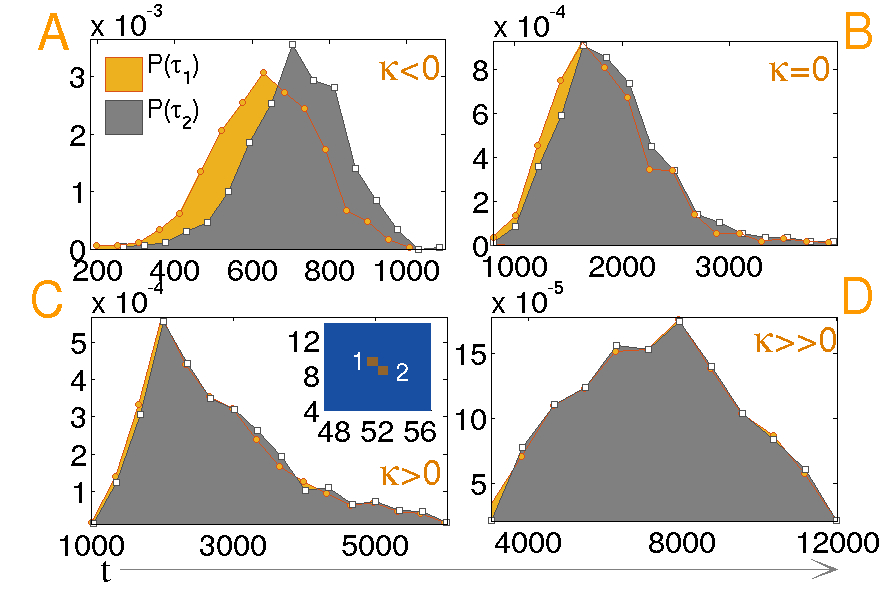}
\caption{\label{fig:pdfIN} (Color Online) (A-D) Normalized probability distribution $P(\tau_1)$ 
of the first nucleation event `1' is displayed along with the probability distribution $P(\tau_2)$ of 
the consecutive event `2' for increasing $\kappa = (-1,0,1,18)$. The spatial proximity of two events 
is shown in the inset of panel(C). $800$ temporal points are sampled to obtain the 
histograms.}
\end{figure} 
 
To evaluate the validity of the CNT, we compute the nucleation rate as a function of the barrier 
height as sketched in fig.(\ref{fig:IN_CNT}). A significant departure from a decaying exponential 
signals a breakdown of the CNT. The CNT deals with the rate of phase change and growth of the
supercritical cluster without accounting for fission and coalescence. Moreover, the occurrence 
of exponential dependency is expected in the Becker-D\"{o}ring limit, {\it i.e.} near the 
coexistence line and for steady state rates. While deformation of the tensorial field due to 
high elastic anisotropy results in the formation of noncircular droplets, the theory can be 
applicable in the $\kappa\to0$ limit where director deformation is negligible and a circular 
shape is retained. In the weak anchoring limit (small $\kappa$), the CNT can still be applied 
if a noncircular shape is incorporated in the standard theory\cite{cutoroijdijk} and the kinetic 
prefactor can be obtained from experimental results. However the CNT has to be supplemented 
to accommodate singularity in ${\bf n}$ in order to make it applicable in the strong anchoring 
limit ($\kappa\gg0$). 
 
Finally to investigate the role of the isotropic medium on the temporal distribution of nucleation 
events, we study the spatiotemporal correlation between the first passage times of two consecutive 
nucleation events. Normalized histograms shown in fig.(\ref{fig:pdfIN}) are sharply peaked for 
$\kappa<0$ and the peak broadens for increasing $\kappa$. Also the distributions are correlated 
in time for $\kappa<0$, although long-ranged elastic interactions are not present in the supercooled 
isotropic film. The reason for this correlation is not clear. The correlation disappears as $\kappa$ 
is increased. Recall that in the isotropic phase, the correlation length is close to few grid spacings, 
so that events separated by more than that are unambiguously recognized as nucleation events. In 
the inset of panel(C), the spatial proximity of two such occurrences are shown. These events are 
temporally uncorrelated in spite of their spatial proximity. For $\kappa=18$ both distributions 
coincide, indicating no memory of consecutive events.

\subsection*{Nucleation in the superheated nematic phase}
\begin{figure}[ht]
\begin{centering}
\includegraphics[width=0.5\textwidth]{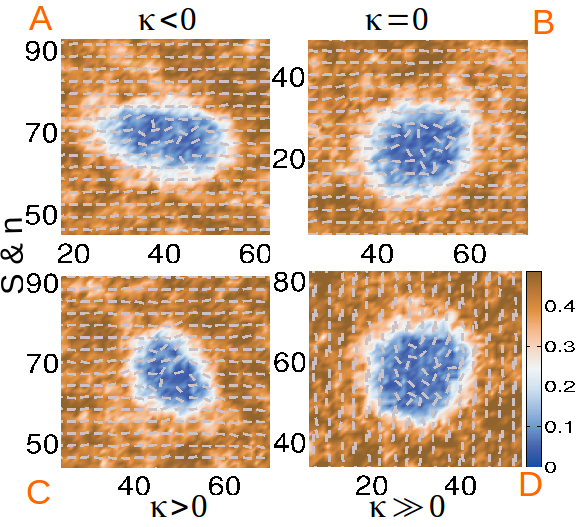}
\par\end{centering}
\centering{}\caption{\label{fig:NImorph} (Color Online) Isotropic droplet structure 
in terms of the uniaxial order parameter and director arrangement in the post-nucleation 
stage of the kinetics, for (A) $\kappa=-1$ at time 
$t=22051\tau$, (B) $\kappa=0$ at time $t=52211\tau$, (C) $\kappa=1$ at time $t=21903\tau$ 
and (D) $\kappa=6$ at time $t=147839\tau$. Scalar 
field values are rendered in false colours.}
\end{figure}
\begin{figure*}[ht]
\includegraphics[width=1.0\textwidth, height=0.7\textwidth]{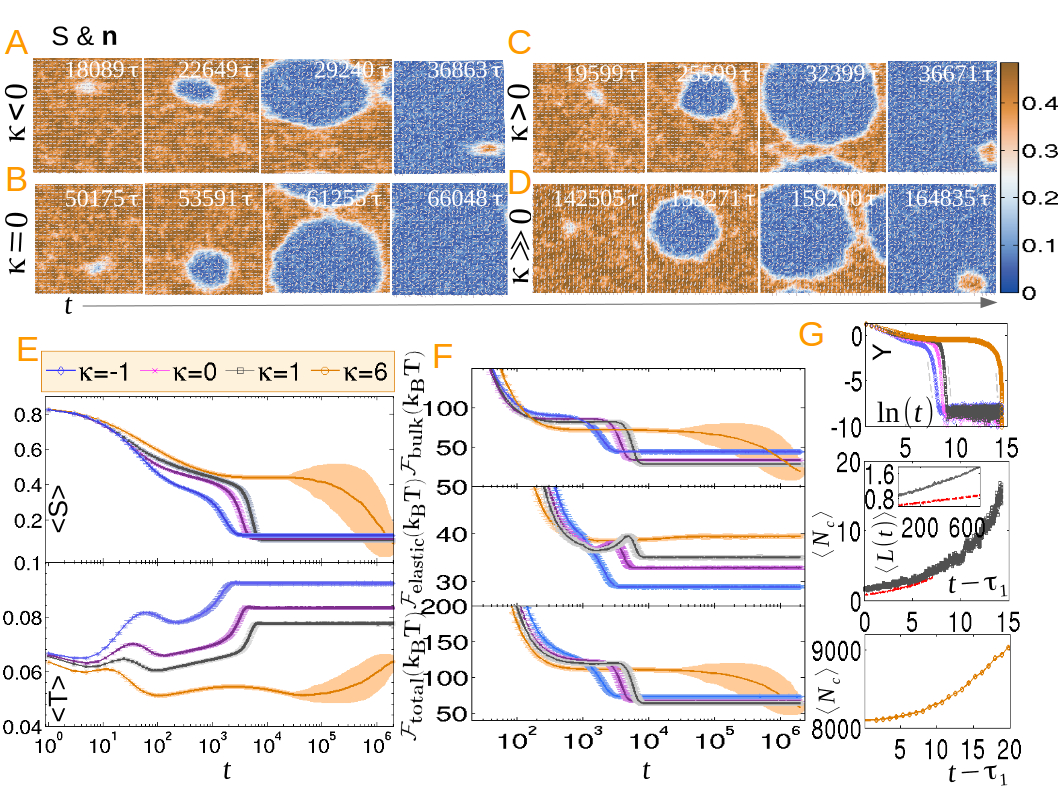}
\caption{\label{fig:FSTevolNI} (Color Online) Panel(A-D): Evolution of uniaxial order and director 
orientation at different stages for higher surface energy and different values of $\kappa$ (see 
Supplementary Animation S3 and S4). Panel (E) displays the evolution of the average uniaxial and 
biaxial order while panel (F) shows bulk, elastic and total energy of the superheated film. The 
upper Panel in (G) presents  fits to the JMAK equation with exponents $m=(2.862,2.952,3.197,4)$ 
for $\kappa=(-1,0,1,6)$ in ascending order. Middle and lower panels in (G) depict evolution of 
$\langle N_c \rangle$ for $\kappa=1$ and $6$. The x-axis corresponds to $(t-\tau_1)\times 10^2$ 
for $\kappa=1$ (middle panel) and $(t-\tau_1)\times10^3$ for $\kappa=6$ (lower panel). Growth of 
tagged cluster size (black line) and average cluster size (red dotted line) are shown in the inset 
of the middle panel in (G). Total $100$ independent realizations are sampled to procure the graphics.}
\end{figure*}
\begin{figure}[ht]
\includegraphics[width=0.5\textwidth, height=0.4\textwidth]{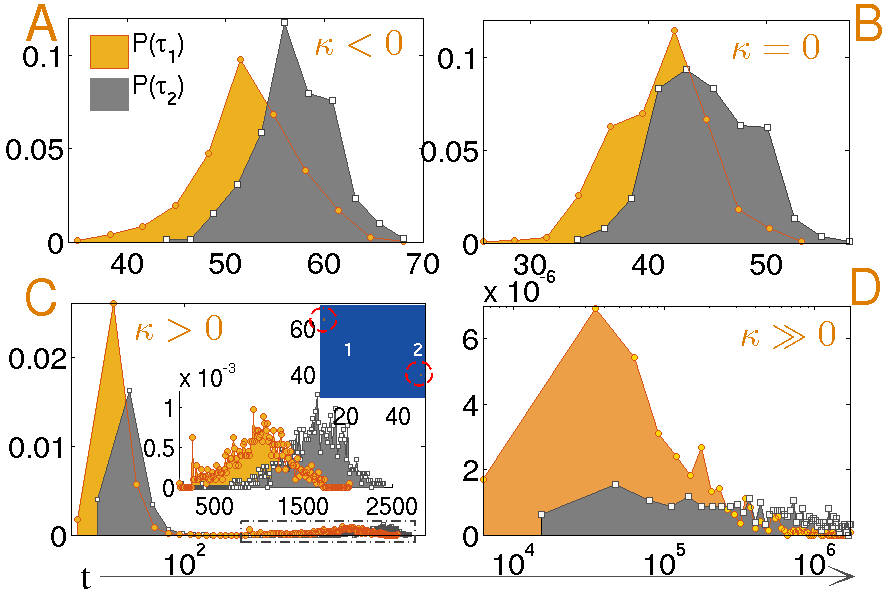}
\caption{\label{fig:pdfNI} (Color Online) (A-D) Normalized probability distributions 
$[P(\tau_1),P(\tau_2)]$ of first and subsequent nucleation events at times $\tau_1$ and $\tau_2$ 
for ascending values of $\kappa = (-1,0,1,6)$. The second peak of the bimodal distribution is 
amplified in the inset of panel (C) where the spatial separation of events `$1$' and `$2$' is 
also portrayed. Histograms are made with $800$ independent points for panel (A-C) while $500$ 
points were sampled to obtain panel (D).}
\end{figure}

The isotropic droplet morphology, evolution of the {\bf Q}-tensor, growth kinetics and temporal 
distribution of nucleation events have also been examined for the case where thermal fluctuations 
nucleate droplets of the isotropic phase in a superheated nematic film. Within feasible computational 
effort, nucleation of isotropic droplets can be obtained only for $\kappa\le6$. Fig.(\ref{fig:NImorph}) 
displays the structure of supercritical droplets at the post-nucleation stage in terms of $S$ 
and ${\bf n}$ for large $L_1$ and different values of $\kappa$. Noncircular droplets nucleate 
for $\kappa\neq0$, while in panel (B) the droplet shape remains nearly circular for $\kappa=0$ 
({\it one elastic constant} approximation). The director distribution is randomized inside the 
droplet, indicating isotropy with no observable biaxiality. Unlike colloidal inclusion in a 
nematic medium\cite{ramnitragpro} or in nematic shells\cite{bates}, homeotropic anchoring at 
the interface by forming defects outside the droplet is not preferred.  

To characterize the evolution process, panels (A-D) of fig.(\ref{fig:FSTevolNI}) portray the 
pre, post, intermediate and late stage structure of $S$ and ${\bf n}$ at a higher surface energy. 
Subcritical droplets form and collapse in the pre-nucleation stage while a supercritical droplet 
nucleates and expands self-similarly in the post-nucleation period. Droplet coalescence converts 
the film into a fluctuating isotropic state at the late stage of the kinetics. In a shallow quench 
where the surface energy and the barrier height is reduced, many small droplets are formed and they 
coalesce with each other. At a late stage, uniform regions of nematic order are squeezed and removed 
from the isotropic film. Rather surprisingly, $\langle S\rangle$ in panel (E) depicts of an unusual 
two-step decay process, while $\langle B_2\rangle$ displays two minima. We interpret this observation
in the following way  (see Supplementary Animation S3). Quenching a uniform nematic medium to 
metastability at a higher temperature induces fluctuations that decrease the scalar order parameter.
The plateau in $\langle S\rangle$ corresponds to its ``quasi-equilibrium'' value in the superheated 
metastable state. For smaller surface energy and a reduced barrier height, the typical size of 
regions of fluctuation-induced melting is comparable to the critical droplet size. Therefore, 
the subcritical droplets do not shrink to zero but persist for sufficient amount of time at the 
pre-nucleation stage, until fluctuations induce the formation of a supercritical droplet. Several 
other mechanisms for the slowing down of the decay of $\langle S\rangle$ may be present, for 
instance (i) fluctuation induced broadening of the zero curvature value of the superheating 
line in fig.(\ref{fig:schematic}), (ii) higher Laplace pressure arising from small droplets, 
(iii) curvature elasticity and capillary force effects. Local heating due to the emission of 
latent heat at the droplet surface can be ignored, while such effects become important at higher 
droplet radius in confined geometry\cite{huifas,chehamshe}. As the minimum in $\langle S\rangle$ 
corresponds to the maximum in $\langle B_2\rangle$, two minima separated by a plateau occurs in 
the lower panel of (E). The post-nucleation droplet growth due to agglomeration is displayed in 
the upper panel of (G) which is characterized by the JMAK equation, with the slope sketched in 
grey dashed lines. Scaling exponents $m>2$ indicate to a breakdown of the CNT description. 

Further support of the two-stage growth process is provided in the evolution of energy as 
highlighted in panel (F). $\mathcal{F}_{elastic}$ slowly decreases after exhibiting two 
overshoots, with the prominent one at a late stage before saturating to the equilibrium value. 
The overshoot corresponds to a maximum in the elastic energy during droplet coalescence. Both 
bulk and total free energies display a plateau where growth remains temporally frozen. As is 
evident, the plateau increases with increasing $\kappa$, indicating that more surface energy 
slows down the formation of supercritical nuclei. These effects are more evident when higher 
surface energy is considered (see Supplementary Animations S3 and S4) where due to increased 
barrier height, the critical radius is large compared to the fluctuation-induced melted 
droplets and subcritical droplets disappear quickly from the film.
 
To quantify the growth process, we explore the evolution of a tagged cluster and the average 
cluster size $\langle N_c\rangle$. Middle and lower panels of (G) display them for $\kappa=1$ 
and $6$. For $\kappa<1$, it was impossible to keep track of single clusters due to very small 
correlation length. Evolution of the average cluster length, shown in the inset of the middle 
panel, is found to follow the tagged cluster dynamics. The growth law is observed to obey 
eq.(\ref{eq:growth}) with a change of an early diffusive to a late stage ballistic growth 
before coalescence. 
 
To examine the role of spatial long-ranged interactions in the first passage times of consecutive 
events, we study the spatiotemporal correlation between the events. Fig.(\ref{fig:pdfNI}) sketches 
the normalized histograms for different $\kappa$. The distribution is sharply peaked within a small 
temporal domain for $\kappa=-1$, and the span of the distribution increases by two orders of 
magnitude with significant broadening as $\kappa$ is increased. The first and second events are 
always correlated due to the long range elastic interaction in the nematic film. The bimodality 
exhibited by the distributions for $\kappa=1$ is surprising. As seen in the amplified plot in the 
inset of panel (C), the second peaks are also correlated. The reason for bimodality can be physically 
understood as the limit in which the size of the regions of fluctuation induced melting becomes less 
than the critical droplet size. Thus the first peak in the histogram in panel (C) results from the 
initial formation of subcritical droplets that disappear in time. However, supercritical droplets 
nucleate at a later stage displayed in the upper inset, with two consecutive events marked as 
`$1$' and `$2$' that are spatially distant but temporally correlated. This should be compared 
with the inset of fig.[\ref{fig:pdfIN}(C)], where although events `$1$' and `$2$' are spatially 
proximate, they are temporally independent. Due to higher surface energy for $\kappa\gg0$, a 
single droplet nucleates and the events in panel (D) are monomodal, but still correlated with 
a much wider temporal distribution compared to that in the weak anchoring limit, shown in panel 
(A-C). 

\section*{Discussion}
\label{sec:4}
We have performed an extensive study of homogeneous nucleation kinetics in a freely suspended 
monolayer of metastable liquid crystalline film using stochastic nematodyamics. In the case of 
a supercooled film in the metastable isotropic phase, we have shown that the presence of a large 
surface interfacial anisotropy quantified by a large value of the parameter $\kappa$ leads to the 
appearance of a noncircular droplet of the nematic phase with an encapsulated hyperbolic hedgehog 
defect and a biaxial interfacial ring as seen in 5CB microdroplets\cite{chehamshe}. Noncircular 
droplets exhibit homogeneous orientation of the director field for smaller values of $\kappa$. 
The growth of the nuclei at small volumes is found to exhibit a polynomial dependence on time. 
The regime of applicability of  classical nucleation theories in the small $\kappa$ limit is 
determined. Also, successive nucleation events are found to be uncorrelated even if they are 
spatially proximate, due to the absence of long ranged elastic interactions in an isotropic 
film. On the other hand, a two step growth process is observed in isotropic droplet nucleation 
in a superheated nematic film. In this case, spatially distant nucleation events are temporally 
correlated due to the long ranged elastic interactions in the nematic film. These findings are 
consistent with available results in three spatial dimensions\cite{bhatam}, but are markedly 
different from the results of studies of confined films where coverslips affect the director 
component in the third direction\cite{lafrmai}.  

The kinetic pathway of ordering from an unstable isotropic phase to a stable nematic phase 
through spinodal decomposition and coarsening in a deterministic GLdG framework has been 
extensively studied\cite{bray,denorlyeom2,zapgold2,subhsoumen,bhmeadd,bhatam} in the past. 
In this case, the development of diffusive domains and late-stage defect pair kinetics (Porod 
law regime) take place at a much faster time scale compared to nucleation kinetics. When a nematic 
film is heated to a temperature above the superheating line, disordered isotropic domain coarsening 
leads to a stable isotropic state\cite{chehamshe}. A comparison of existing results for late-stage 
domain growth in these cases with those obtained from the stochastic GLdG framework considered here 
is outside the scope of the present study. Also, electrokinetic\cite{olavmool} and flexoelectric 
effects\cite{bardur} as well as coupling of the orientation tensor to a hydrodynamic flow field\cite{beredw} 
for a thermal system can be considered in the future. Other choices for describing biaxial order\cite{bezan}
may be explored in future investigations. Experimental verification of the results reported here 
would be welcome, although avoiding heterogeneous nucleation when sampling rare events in a narrow 
temperature window is a challenging task. 

\section*{Methods}

{\bf Numerical integration of stochastic GLdG equation.} 
A two dimensional monolayer of nematogenic material is considered, where orientation in three 
Cartesian directions is retained, but spatial variations are restricted to a plain. The {\bf 
Q}-tensor equation is solved on a regular square lattice with periodic boundary condition to 
neglect confinement effects. A direct numerical integration is forbidden as using similarity 
transformation, any symmetric traceless tensor cannot be diagonalized at every grid point. By 
utilizing a property that the tensor can be expanded in a basis of five $3\times3$ matrices 
${\bf T}$\cite{kawehe}, a legitimate way is to project the equations as ${\bf Q} = \sum_i a_i {\bf T}_i$ 
and ${\pmb \xi} = \sum_i \zeta_i {\bf T}_i$ $(i=1,\ldots,5)$, so as to contain the dynamics 
in the basis coefficients $a_i({\bf x},t)$ and $\zeta_i({\bf x},t)$\cite{bhmeads}. Major 
advantage is gained in constructing the symmetrized detraced noise ${\pmb \xi}$ with five 
$\zeta_i$, that corresponds to zero mean unit variance independent Gaussian white noise 
processes. This thus validates discrete FDT spectrum in all Fourier modes and we obtain 
reasonable agreement in static and dynamic correlations of ${\bf Q}$ with analytic formula 
both in isotropic and nematic phase \cite{bhmeads}. Eq.(\ref{eq:Qdynamics}) in the basis 
coefficients takes the form 
\begin{eqnarray}
\label{eq:adynamics}
&& \partial_{t}a_{i} = - \Gamma \big[(A + C \mathrm{Tr}Q^{2})a_{i} + BT_{\alpha\beta}^{i}
\stl{{Q_{\alpha\beta}^{2}}} - L_{1}\partial_\alpha^{2}a_{i} - \nonumber \\ 
&& L_{2} \stl{T^{i}_{\alpha\beta}T^{j}_{\beta\gamma}\partial_{\alpha}
\partial_{\gamma}a_{j}} - L_3 \big\{Q_{\alpha\beta}\partial_{\alpha}\partial_{\beta}a_{i} - 
T^{i}_{\alpha\beta}\partial_{\alpha}a_{j}\partial_{\beta}a_{j} \big\}\big] \nonumber \\ 
&& + \zeta_i
\end{eqnarray}
where $\langle \zeta_{i}({\bf x}, t)\zeta_{j}({\bf x}^{\prime}, t^{\prime}) \rangle =
2k_BT\Gamma \delta_{ij}\delta({\bf x - x^{\prime}})\delta(t - t^{\prime})$.

Laplacian and mixed derivatives are spatially discretized as $\partial_m^2 a(m,n) = 
[a(m+1,n) + a(m-1,n) - 2a(m,n)]/(\Delta m)^2, \partial_m\partial_n a(m,n) = [a(m+1,n+1) - 
a(m+1, n-1) - a(m-1,n+1) + a(m-1, n-1)]/{4\Delta m\Delta n}$, where $m,n$ denote Cartesian 
indices. We adopt second order accurate stochastic method of lines (SMOL) integrator for 
explicit temporal update\cite{wilkie}. SMOL semi-discretization scheme develops on 
discretizing spatial part of partial differential equations to yield ordinary time-dependent 
equations, which are integrated on unstructured grid maintaining accuracy, stability 
and computational overload. 

The distortion free energy, length and time are resolved by transforming the deterministic 
part of eq.(\ref{eq:Qdynamics}) in non-dimensionalized form to obtain $l^{*}_{(\kappa>0)} = 
5\sqrt{18CL_1(1 + 2\{\kappa+\Theta\}/3)}/3B, l^{*}_{(\kappa<0)} = 5\sqrt{18CL_1(1 + \{\kappa+
\Theta\}/6)}/3B, \mathcal{F}^{*} = 9C S_c^4/16, \; t^{*} = \Gamma F^{*}/S_c^2, \; \Upsilon^{*}_{(\kappa \neq 0)} = 
\mathcal{F}^{*}l^{*2}_{(\kappa \neq 0)}$, where $l^{*}_{(\kappa \neq 0)},\mathcal{F}^{*},
t^{*}$ and $\Upsilon^{*}_{(\kappa\neq0)}$ are non-dimensional length, bulk energy, time 
and surface energy. Dimensional quantities for example, correlation length and relaxation 
time can be computed as $\lambda = \sqrt{32}l^{*}_{(\kappa \neq 0)}/{3(1 + \sqrt{1 - 
24AC/B^2})}, \;\tau = t^{*} \Delta t$. To avoid numerical artifact, $t^{*}\ll 1$ and 
$\lambda \gg \Delta x$ are strictly maintained. Also $\Upsilon^{*}_{(\kappa\neq0)}\gg 
k_BT$ is ensured to avoid the medium to attain the stable phase in one computational step.

{\bf Cluster labelling procedure.} To sample nucleation clusters, we record results on 
every computational step within a time window within which the cluster eventuates. We apply 
Hoshen Kopelman (HK76) algorithm \cite{hk76} to label connected clusters on the grid which 
are above (below) certain threshold. To identify nematic nuclei, we choose threshold value 
at $70\%$ of $S_{eq}$ and implement periodicity in both directions to overcome double 
counting of connected clusters through periodic boundaries. In case of isotropic nucleation, 
the algorithm performs reversely and we choose the threshold value at $30\%$ of $S_{eq}$. 
The algorithm particularly finds usefulness in counting the total number of grid points 
pertaining to a tagged cluster that temporally amplifies as the cluster swells. Thus a 
length scale can be simply extracted to quantify growth law, without computing the length 
scale from direct correlation functions\cite{bhmeadd} that also captures Porod law 
scaling of defect annealing kinetics after droplet coalescence. 

\bibliography{references}

\section*{Acknowledgements}
We thank R. Adhikari for suggesting the problem and thank including him G.I. Menon and C. Dasgupta 
for a critical reading of the manuscript. We extend thanking D. Frenkel, S. Ramaswamy, S. Dhara, 
R. Pratibha, V.A. Raghunathan for stimulating discussions, A. Laskar for suggesting HK76 algorithm 
and reviewers for informative remarks. We acknowledge funding through the DST-INSPIRE program and 
Matscience Chennai for a short term visit. 

\section*{Additional information}
Supplementary information in conjunction to this article at http://www.nature.com/scientificreports \\

Competing financial interests: The author declares no competing financial interests.

%%%%%%%%%% Merge with supplemental materials %%%%%%%%%%
\newpage
\begin{widetext}
\begin{center}
\textbf{\large Supplemental Information}
\end{center}
\end{widetext}

\begin{center}
{\bf Breakdown of {\it de Gennes ansatz} on curved surface interface} \\
\end{center}

In a principal frame, the diagonal components of $\bf{Q}$ are written as, $Q_{xx}= -(S+T)/2, 
Q_{yy} = -(S-T)/2, Q_{zz}=0$. The matrix can be transformed to a fixed frame of reference by
rotation with pitch angle $\theta$ and yaw angle $\phi$ to obtain,
$Q_{xx} =-\big\{(S+T)\cos^2\phi\cos^2\theta + (-S+T)sin^2\phi cos^2\theta + 2S sin^2\theta\big\}/2, 
Q_{xy} = -Tsin(2\phi)cos\theta/2, Q_{xz} = \big\{3S + Tcos(2\phi)\big\}sin(2\theta)/4, 
Q_{yy} = \big\{-S + T cos(2\phi)\big\}/2, Q_{yz} = Tsin(2\phi) sin\theta/2, Q_{zz} = 
\big[2S - 2Tcos(2\phi) + Tcos\big\{2(\phi-\theta)\big\} + 6Scos(2\theta) + Tcos\big\{2(\phi + 
\theta)\big\}\big]/8.$ Inhomogeneities of $S, T, \theta$ and $\phi$ are encoded in
$\mathcal{F}_{elastic}$, whose minimization for different $\kappa$ decides stable director
anchoring at surface interface.

In absence of thermal fluctuations (${\pmb\partial}\theta, {\pmb\partial}\phi=0$) for a
planar I-N interface along {\it${z}$}-direction where the director is confined to a plain
($\phi=0$), the anisotropic elastic energy takes the form
\begin{eqnarray}
\mathcal{F}_{anelastic} &=& \kappa 
\big[sin^2(2\theta)(3\partial_z S+\partial_z T)^2 + \big\{1+3cos(2\theta) \times \nonumber \\ 
&& \partial_zS - 2sin^2\theta\partial_z T\big\}^2\big]/32.
\end{eqnarray}
Thus, free energy is lowered for homeotropic anchoring ($\theta=0$) for $\kappa<0$ and
planar anchoring ($\theta=\pi/2$) for $\kappa>0$, in par with de Gennes argument \cite{degennes}.
However for a curved interface,
\begin{widetext}
\begin{eqnarray}
\mathcal{F}_{anelastic}^{(\theta=0)} &=& \kappa\big[(\partial_z S)^2 + \big\{\partial_yS/2 
+ sin(2\phi)\partial_x T/2 + (1/2 - cos\phi)\partial_yT\big\}^2 + \big\{\partial_xS + 
cos(2\phi)\partial_xT + sin(2\phi)\partial_yT\big\}^2/4\big]/2, \nonumber\\
\mathcal{F}_{anelastic}^{(\theta=\pi/2)} &=& \kappa\big[(\partial_x S)^2 + \big\{-
\partial_zS + sin(2\phi)\partial_y T - cos(2\phi)\partial_zT\big\}^2/4 + \big\{-\partial_yS 
+ (-1/2 + cos\phi)\partial_yT + sin(2\phi)\times \nonumber \\ 
&& \partial_zT/2\big\}^2\big]/2,
\end{eqnarray}
\end{widetext}
which can be further reduced in a quasi two-dimensional geometry by taking $\partial_zS,
\partial_zT=0$. Depending on the sign of $\kappa$ and according to the competing values
of the gradients in $S,T$ and $\phi$, the film decides the favoured anchoring. Accounting
to thermal fluctuations $({\pmb\partial}\theta,{\pmb\partial}\phi\neq0)$, director anchoring
at the droplet surface interface is not intuitive.

\begin{center}
{\bf Time-lapse animations \& Captions} \\
\end{center}

{\bf Animation S1} : {\it Circular nematic droplets for $\kappa=0$ and lower surface energy.} \\

Description: The animation sequentially portrays evolution of (a) $S$ \& ${\bf n}$, (b) $T$ \&
${\bf l}$ and (c) Schlieren texture in {\it one elastic approximation} and for small values of
elastic constant $L_1$. Nucleation of circular nematic bubbles with uniform director field is
observed, that amplify in size to coalesce with other droplets. Note that many droplets are
formed as in shallow quench, and droplet coalescence resulted into defects of half integer
charge due to lower surface energy. For higher surface energy (or larger $L_1$), only few
droplets are nucleated whose coalescence does not generate defects (not shown). Almost no
notable change in $T$ and ${\bf l}$ fields is seen in the process. Finally, schlieren
texture depicts the uniformity of director field within the droplets. \\

{\bf Animation S2} : {\it Noncircular nematic droplets with encapsulated hyperbolic hendgehog
defects for $\kappa\gg0$ and higher surface energy.} \\

Description: The animation sequentially portrays evolution of (a) $S$ \& ${\bf n}$, (b) $T$ \&
${\bf l}$ and (c) Schlieren texture in strong anchoring limit and for higher $L_1$. Nucleation of
noncircular nematic bubbles with encapsulated defect is observed, that amplify in size to
coalesce with other droplets. The formation of biaxial ring at the droplet interface with
hyperbolic hedgehog defect structure is also observed in $T$ and ${\bf l}$ fields. Finally,
schlieren texture depict the $4$-brush geometry, that persist at very late stage of the
kinetics. \\

{\bf Animation S3} : {\it Double occurance of noncircular isotropic droplets for $\kappa=1$ and
lower surface energy.} \\

Description: The animation sequentially portrays evolution of (a) $S$ \& ${\bf n}$ and (b) Schlieren
texture in weak anchoring limit and for lower $L_1$. Almost no notable change in $T$ and ${\bf l}$
field is seen and thus omitted from the animation. Nucleation of noncircular isotropic bubbles
without any defect in the bulk nematic film is observed. Note the double occurance of droplets,
resulting to bimodality in the probability distribution of nucleation events. Nucleated droplets
in later stage amplify in size to coalesce with other droplets and span the system size to form
isotropic phase. Schlieren textures depict the nonuniformity of the director field more
transparently. \\

{\bf Animation S4} : {\it Noncircular isotropic droplets for $\kappa=6$ and higher surface energy.} \\

Description: The animation sequentially portrays evolution of (a) $S$ \& ${\bf n}$ and (b) Schlieren
texture in strong anchoring limit and for higher $L_1$. Nucleation of noncircular isotropic bubbles
is observed that amplify in size to coalesce with other droplets. Schlieren textures support the
nonuniformity of director field within the droplets and absence of defects in the nematic environment,
as well as in the squeezing nematic domains at the late stage of the kinetics.

\end{document}